\documentclass[prb,twocolumn,showpacs,amsmath,amssymb,floatfix,superscriptaddress]{revtex4-2}

\usepackage{import}

\usepackage{color,graphics,epsfig,rotating}
\usepackage{graphicx}
\usepackage{dcolumn}
\usepackage{bm}

\usepackage{placeins}
\usepackage{mathrsfs}
\usepackage{lmodern}
\usepackage{notoccite}

\usepackage[breaklinks=true]{hyperref}
\usepackage{breakcites}
\usepackage{cleveref}
\usepackage{amsmath}
\usepackage{grffile}
\usepackage{xspace}
\usepackage{xifthen}
\usepackage{soul}
\usepackage{placeins} 

\usepackage{natbib}

\usepackage[normalem]{ulem}

\newcommand{\pd}{\partial}

\newcommand{\xc}{\mathrm{xc}}
\newcommand{\eq}{\mathrm{eq}}
\newcommand{\lda}{\textsc{lda}}

\newcommand{\Av}[1]{\left\langle{#1}\right\rangle}

\begin{document}

\title{Accurate \textit{ab initio} modeling of solid solution strengthening in high entropy alloys}

\author{Franco Moitzi}
\affiliation{Materials Center Leoben Forschung GmbH, Roseggerstra{\ss}e 12, A-8700 Leoben, Austria}
\author{Lorenz Romaner}
\affiliation{Chair of Physical Metallurgy and Metallic Materials, Department of Materials Science, 
             University of Leoben, Roseggerstra{\ss}e 12, A-8700 Leoben, Austria}
\author{Andrei V. Ruban}
\affiliation{Materials Center Leoben Forschung GmbH, Roseggerstra{\ss}e 12, A-8700 Leoben, Austria}
\affiliation{Department of Materials Science and Engineering, Royal
         Institute of Technology, 10044 Stockholm, Sweden}
\author{Oleg E. Peil}
\affiliation{Materials Center Leoben Forschung GmbH, Roseggerstra{\ss}e 12, A-8700 Leoben, Austria}


\received{29 July 2022}
\accepted{29 September 2022}
\published{24 October 2022}

\begin{abstract}
High entropy alloys (HEA) represent a class of materials with promising properties, such as high strength and ductility, radiation damage tolerance, etc. At the same time, a combinatorially large variety of compositions and a complex structure render them quite hard to study using conventional methods. In this work, we present a computationally efficient methodology based on 
\textit{ab initio} calculations within the coherent potential approximation. To make the methodology predictive, we apply an exchange-correlation correction to the equation of state and take into account thermal effects on the magnetic state and the equilibrium volume. The approach shows good agreement with available experimental data on bulk properties of solid solutions. As a particular case, the workflow is applied to a series of iron-group HEA to investigate their solid solution strengthening within a parameter-free model based on the effective medium representation of an alloy. The results reveal intricate interactions between alloy components, which we analyze by means of a simple model of local bonding. Thanks to its computational efficiency, the methodology can be used as a basis for an adaptive learning workflow for optimal design of HEA.
\end{abstract}

\maketitle

%
%

%
%
\section{Introduction}

Concentrated solid solutions with multiple components, also known as high
entropy alloys (HEA), have attracted a lot of attention because of their
exceptional combination of yield strength and ductility \cite{Oh2019,George2019,Cantor2014,Steurer2020,Schuh2015}. Although 
the highest strength is usually achieved in multiphase
alloys, HEAs are characterized by a high yield strength even in the form of
single-phase solid solutions \cite{Cantor2014}. Especially interesting
in this respect are face-centered-cubic (fcc) HEAs, where a high ductility
can coexist with high strength beyond the usual tradeoff \cite{Li2016}.
The high strength in these systems is mainly associated with the
interaction of dislocations with solutes, which impedes the motion
of dislocations, resulting in the solid solution strengthening (SSS).

Direct modeling of dislocation motion in multi-component alloys is currently
possible only within semi-empirical approaches such as classical
molecular dynamics \cite{Yamakov2001,Chang2002,Olmsted2006}, which
provide a qualitative picture.
At the same time, comparable \textit{ab initio}
simulations of dislocations are not feasible, and one has to combine
them with phenomenological models, reducing the complex problem
of a dislocation interacting with alloy components to a set of parameters
(e.g., effective atomic volumes, elastic moduli, the stacking fault energy, etc.)
that can be obtained by first-principles calculations.

Recently, Varvenne and Curtin (VC) have proposed a generalized model
that does not rely on any assumptions about the concentration of components 
and it is specifically designed to treat HEA systems \cite{Varvenne2016,Varvenne2017a}.
Similar to earlier dilute-limit models \cite{Fleischer1963,Labusch1970,Nabarro1977},
the central mechanism of the VC model
is the occurrence of energy barriers for dislocation motion due to 
local effective volume fluctuations caused by alloy components, which
increases the stress for the onset of dislocation glide and hence the yield strength \cite{Varvenne2016}.
The system is treated as an effective alloy matrix with solutes
interacting with the matrix by means of stress fields caused by the atomic size
mismatch between a solute and the effective matrix element. 
Thanks to the low Peierls stress in face-centered-cubic (fcc) metals,
the yield stress is then expressed solely
in terms of the average misfit volume of components and linear elastic properties
of the alloy. An important feature of the VC model is that it
suggests that the strengthening effect does not directly
depend on the number of components and is not necessarily maximized by the equimolar
composition often used in experimental work \cite{Basu2020}. 
Instead, the largest yield stress is achieved by maximizing the
average mean-square misfit volume, while minimizing 
a potential negative impact on the elastic properties.
This is consistent with experimental evidence that nonequimolar
compositions can show significantly improved mechanical properties as
compared to alloys with equimolar ratios of principal elements
\cite{Yao2014,Pradeep2015}.

The VC model has been successfully applied to a number of systems
\cite{Varvenne2017b,Tehranchi2017,Laplanche2018a,Bracq2019,Yin2020a,Yin2020b}.
However, in most of the cases the model parameters have been determined 
or estimated from experimental
data, which limits the applicability and predictive power of the model.
On the other hand, these parameters could be calculated \textit{ab initio}
using the density functional theory (DFT), but
the accuracy of the calculation is strongly affected by the general
error that most widely used (semi)local exchange-correlation (XC) functionals
produce for the equilibrium volume of an alloy or an element in its ground
state. This leads to an overall discrepancy both in the average molar
volume of the alloy and in the misfit volumes of constituent elements
\cite{Yin2020a}.
The situation becomes even more involved in the case of HEA based on
the iron group of 3d metals, where finite-temperature magnetism plays an
important role for relevant temperatures of interest, usually ranging
from room temperature to 1000~K and above.

In this work, we address the above-mentioned limitations of DFT-based modeling
by proposing a computationally inexpensive methodology based on the coherent
potential approximation (CPA) \cite{Soven1967,Velicky1968},
which offers a very consistent way of describing
properties of the effective alloy medium underlying the VC model.

Within this approach, the equilibrium volumes are adjusted 
using element-specific corrections, which, in turn,
improve other equilibrium properties. Our methodology also takes into account 
finite-temperature effects mediated
by magnetic and phonon degrees of freedom, which are especially important
for describing the temperature dependence of SSS above room temperature.
We create a fully automatized workflow for calculating all necessary
quantities for estimating SSS and demonstrate it on several well-studied 
3d transition metal HEAs, 
which are well known for their complex magnetic behavior.

%
%

\section{Methods}
\label{sec:methods}

\subsection{Exchange-correlation pressure correction}

The use of local [e.g., local density approximation (LDA)] or semilocal
[e.g., generalized gradient approximation (GGA)] exchange-correlation
functionals within DFT is known to lead to systematic errors in
equilibrium properties of solids.
Multiple attempts to improve XC functionals have been
made \cite{Perdew1999,Tao2003,Zhao2006,Perdew2008,Sun2015}
but due to general limitations of semilocal density functionals
\cite{Becke1999} these implementations often improve the accuracy 
for only certain classes of solids, while failing to do so for others.

One can distinguish two types of errors: (1) the error in
the equilibrium volume (or lattice constant), and
(2) the error in the value of a target quantity
at the exact experimental volume. The motivation for this distinction
comes from the empirical observation that many linear-response properties
can be reproduced reasonably well even by the local density approximation
without gradient corrections, provided that a calculation is performed
at the experimental equilibrium volume \cite{Wu2006}.
Whereas type-2 errors are intrinsic to an XC functional and cannot be
remedied without reconsidering the XC functional itself,
type-1 errors can be estimated and eliminated since
the equilibrium volume of practically all pure elemental materials
is known experimentally, and we only need to find a way how to use
this information to correct for the volume error of a compound. 
A naive interpolation of volumes or lattice constants (Vegard's law)
from experimental values for pure elements will definitely give
poor results because the interaction between different elements in a compound
is completely ignored in this case. Another approach is to apply linear
interpolation only to the deviations between DFT and experimental results
\cite{Razumovskiy2019}. This way one can get much better results for solid
solutions of similar elements but this methodology is not
well justified in cases when the ground state structure of components
is different from the structure of an alloy.
For example, it is not clear how one
could correct for the error in determining the equilibrium volume of
a hypothetical zero-temperature fcc structure of Fe (which can serve
as a reference for Fe-based fcc alloys) from the error
in the actual ground-state body-centered-cubic (bcc) structure.

A different approach was put forward by van de Walle and Ceder \cite{VandeWalle1999}
who argued that there is an intrinsic source of errors in the traditional
XC functionals (LDA, GGA, meta-GGA) related to their
semilocal nature. Based on their systematic study of the error in the
equilibrium volume of multiple ordered compounds they concluded that this
error can be largely attributed to a non-local
contribution responsible for a significant modification of
the exchange-correlation hole of interstitial (valence) electrons
in the region predominantly occupied by highly localized core electrons. 

Under certain rather general assumptions about the form of this
non-local contribution, they concluded that its effect on the
total energy can, to a first approximation, be taken into account by
adding to the local density functional another term that is linear in volume.
This results in an additional \textit{pressure}
correction associated with each atom in a compound. 

Importantly, this correction turns out to be also linear in concentration,
implying that most of the error in the calculation
of the equation of state can be eliminated by introducing
an XC pressure correction (XPC), $P_{\xc}$,
given by a sum of individual element-specific contributions,
$P_{\xc} = \sum_{i} m_{i} P^{(i)}_{\xc}$, for elements $i$ of a unit cell,
with $m_{i}$ denoting the element multiplicity. The linear dependence of $P_{\xc}$ on concentrations
makes the above ansatz readily applicable to disordered alloys,
where the XC pressure correction will be given
by $\Av{P_{\xc}} = \sum_{i} c_{i} P^{(i)}_{\xc,i}$ with $c_{i}$
being the atomic fractions of alloy components $i$.
Moreover, the simple form of the correction makes it compatible
with the CPA for disordered systems
\cite{Soven1967,Velicky1968},
suggesting a simple and efficient approach for accurate evaluation of
the equilibrium volumes and other properties of solid solutions. 

Given the known experimental zero-temperature equilibrium volume,
$V^{(i)}_{\eq}$ (corrected for zero-point vibrations),
of element $i$, the parameter $P^{(i)}_{\xc}$ correcting
a given XC functional -- say, LDA -- can be determined from
\begin{align}
P^{(i)}_{\xc} = -P^{(i)}_{\lda}(V^{(i)}_{\eq}),
\end{align}
where $P^{(i)}_{\lda}(V)$ is the LDA (pressure) equation of state for
element $i$.
This ensures that the corrected pressure, $P^{(i)}_{\lda}(V) + P^{(i)}_{\xc}$,
equals zero at the exact experimental volume. 

Once the average parameter $\Av{P_{\xc}}$ is known, the corrected equation
of state reads
\begin{align}
\overline{E}(V) = \Delta E_{\xc} + E_{\lda}(V)  - \Av{P_{\xc}} V,
\end{align}
where $E_{\lda}(V)$ is the equation of state within the LDA functional
and $\Delta E_{\xc}$ is a constant correction term.
This term, $\Delta E_{\xc}$, can be
estimated in the same fashion as the pressure
from the difference between the theoretical and experimental
cohesive energies. However, it cancels out for most of the
properties that preserve species balance and we, therefore, do not consider
it in this work. The correction for other semilocal functionals, such as Perdew-Burke-Ernzenhof
(PBE) \cite{pbe96} or Perdew-Wang (PW91) \cite{pw91}, is formulated
in a similar way.

In this, we use the exact muffin-tin orbital (EMTO)
\cite{Vitos2001a} code
(Lyngby version \cite{Ruban2016}) implementing a Green's function
based DFT methodology combined with CPA to perform total energy calculations.
Screened Coulomb interactions in the CPA are obtained by
the use of the locally self-consistent Green’s function 
technique \cite{Abrikosov1997} implemented also within the EMTO method \cite{Peil2012}.
The total energy is obtained within the full charge density
formalism \cite{Vitos1997}, making the results comparable
to those from full potential codes.
The paramagnetic state of metallic alloys is described using the
disordered local moment (DLM) approach \cite{Gyorffy1985}.

\subsection{Finite-temperature effects}\label{sec:finite-T}

Most of the experimental results for alloys are obtained at room temperature.
Moreover, for many technological alloys it is vital to understand
their behavior at even higher temperatures.  We, therefore, need to extend
calculations of the equation of state to finite temperatures.
For the ambient pressure we replace the total energy with the Helmholtz free energy,
\begin{align}
F(V, T) = F_{\mathrm{el}}(V, T) + F_{\mathrm{ph}}(V, T)
\end{align}
where we consider two main contributions: electronic, $F_{\mathrm{el}}(V, T)$,
containing the XC pressure correction as well as eventual high-temperature magnetic fluctuations,
and phonon, $F_{\mathrm{ph}}(V, T)$.
Such a decomposition is possible because phonon degrees of freedom evolve
on a much longer time scale than the electronic ones. 

The electronic part of the free energy can, in turn, be split
into the one-electron part, $F_{\mathrm{one}}(V, T)$, and the magnetic part,
$F_{\mathrm{mag}}(V, T)$.
The one-electron entropy is calculated using the usual Sommerfeld formula
with energy contour integration weighted by the Fermi function. 
Unlike the one-electron part, the magnetic contribution to
the free energy, $F_{\mathrm{mag}}(V, T)$, is more difficult to tackle
in a consistent manner. Although DLM is supposed to describe the
paramagnetic state at high temperatures, in its original form it works
only for elements (Fe and Mn) which preserve the localized character of
spin magnetic moments in the DLM state. To get the correct magnetic behavior
for elements with primarily itinerant character of magnetism (Cr, Ni, Co)
one has to take into account temperature-induced longitudinal spin fluctuations (LSF).
In this work, we use a semiclassical model presented earlier in
Refs.~\onlinecite{Ruban2013,Ruban2016}. Within this model, we perform DLM
calculations and take into account LSF by adding
an entropic contribution, $-T S_{\textrm{mag}}[M]$, to the
electronic free energy and to the self-consistent potential. 
Here, $M$ is the magnetic moment of a specific component
obtained self-consistently within the DFT cycle. For elements with primarily itinerant
magnetic character, the entropy has a form,
$S_{\textrm{mag}}[M] = a \log M$, where coefficient $a$ is element-specific
and furthermore depends on the magnetic behavior of
an element in a particular alloy system. The coefficient can be chosen
based on a series of fixed-spin-moment calculations for the element in a
given alloy \cite{Ruban2016}. For elements with more localized behavior,
i.e., Fe (at large volumes) and Mn, the entropy term is taken to be equal to
$S_{\textrm{mag}}[M] = \log (1 + M)$. We note that this model of LSF has
been producing very consistent results for alloys based on iron-group metals,
including Fe-based fcc solid solutions, which is generally a rather difficult
case for modeling \cite{Razumovskiy2016a,Razumovskiy2019,Kabliman2011,Dong2017}.

The phonon free energy is calculated within the Debye-Gr\"uneisen model
\cite{Moruzzi1988,Korzhavyi1994}, with the parameters
(the bulk modulus, the Gr\"uneisen constant) derived from the equation of state
calculated at the respective temperature. This way, basic coupling between
magnetic and phonon degrees of freedom related to the volume dependence of
the magnetic moment (the invar/anti-invar effect) is taken into account.

\begin{table}[h!]
\centering
\caption{Alloys used for testing the pressure correction on equilibrium volume and bulk modulus.}
    \begin{tabular}[t]{|l|c|c|c|c|}
    \hline
    Composition  & Magnetic state  & T (K) & Structure & Ref. \\ \hline
    Ni           &   Paramagnetic  &       700       &  fcc      & \cite{Hwang1972,Ledbetter1973} \\ \hline
    NiCo         &   Paramagnetic  &       300       &  fcc      & \cite{Gao2016} \\ \hline      
    NiCoCr       &   Paramagnetic  &       300       &  fcc      & \cite{Gao2016} \\ \hline      
    NiCoCrMnFe   &   Paramagnetic  &       300       &  fcc      & \cite{Haglund2015,Gao2016} \\ \hline      
    NiCoFe       &   Paramagnetic  &       300       &  fcc      & \cite{Gao2016}\\ \hline      
    NiCoCrFe     &   Paramagnetic  &       300       &  fcc      & \cite{Gao2016} \\ \hline      
    Fe0.2Cr0.8   &   Paramagnetic  &       300       &  bcc      & \cite{Speich1972,Zhang2013} \\ \hline      
    Fe0.88Co0.8  &   Ferromagnetic &       300       &  bcc      & \cite{Speich1972} \\ \hline      
    W0.3Cr0.7    &   Nonmagnetic  &       300       &  bcc      & \cite{Landa2021} \\ \hline
    W0.8Cr0.2    &   Nonmagnetic  &       300       &  bcc      & \cite{Landa2021} \\ \hline
    Fe0.94Re0.06 &   Paramagnetic  &       300       &  bcc      & \cite{Speich1972} \\ \hline
    \end{tabular}

\label{tab:xc_samples}
\end{table}

The described methodology was benchmarked by applying it
to a series of alloys for which reliable
experimental data on the equilibrium volume and elastic properties are
available (Table \ref{tab:xc_samples}). For reference calculations of magnetic 3$d$ metals their respective ground
state magnetic structures were considered. The spin-density wave magnetic
state of Cr was approximated by a collinear anti-ferromagnetic state
of [001] type. The magnetic state of $\alpha$-Mn has been
approximated by a collinear anti-ferromagnetic state structure \cite{Sliwko1994}.

\begin{figure}[!ht]
   \centering
   \includegraphics{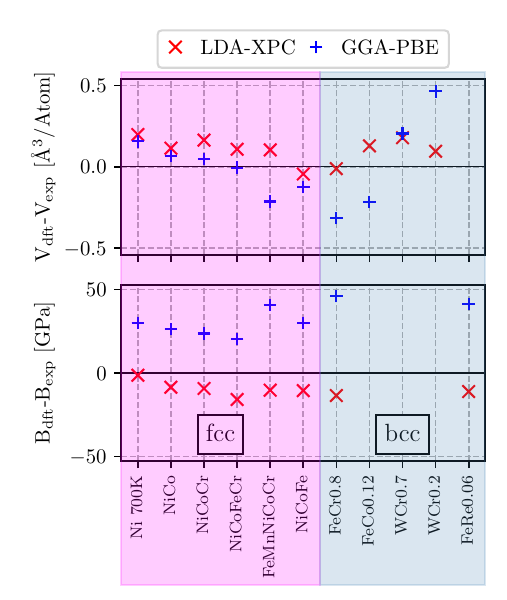}
   \caption{
   Comparison of the experimental and calculated
   atomic volume and bulk modulus for disordered paramagnetic and nonmagnetic alloys. ($\times$ red): LDA with XC pressure-correction (LDA-XPC). ($+$ blue): PBE. 
   Phonon contributions to the thermal expansion are included in all cases. Magnetic entropy is considered for the paramagnetic alloys.
   }
   \label{fig:pressure_correction_procedure}
\end{figure}

In the calculations with XC pressure correction we choose LDA as the reference functional.
This choice is motivated, on one hand, by better magnetic moments that
this functional tends to produce
at the experimental volume. On the other hand, the relatively large error in
the equilibrium volume is anyway corrected by the pressure correction. 
In contrast to LDA, the most commonly used gradient approximation, PBE,
is known to overestimate magnetic effects \cite{Wu2006,Park2015} and it is, hence,
less preferable for systems where magnetism is crucial. 
The results of the benchmark calculations with LDA-XPC compared to
a fully \textit{ab initio} PBE and to experimental data \cite{Speich1972,Zhang2013,Huang2019,Lucas2012,Landa2021,Haglund2015,Brechtl2022,Gao2016,Ledbetter1973}
are displayed in Fig.~\ref{fig:pressure_correction_procedure}. 
Pure LDA results were omitted from the figure due to their largely underestimated
equilibrium volumes compared to LDA-XPC and PBE. As one can see, equilibrium volumes obtained with LDA-XPC calculations
are in better agreement with experiment for most of the cases except for three
systems where PBE performs slightly better but both methods give
a rather small error in these cases. At the same time, when it comes to the bulk modulus, LDA-XPC always performs
significantly better than PBE, reducing the relative error from more than 16~\%
to less than 10~\%.

\subsection{Solid solution strengthening}

The VC model predicts the critical resolved shear stress (CRSS) at any 
temperature and strain-rate for any given alloy composition, with CRSS 
defined as a stress required to initiate slip in a perfect single crystal. 
It can be considered as a material-specific and temperature-dependent constant,
independent of the microstructure of a sample and the specific measurement method. 
In HEAs considered in this work, the CRSS is mostly determined by SSS,
since the Peierls stress can be disregarded in fcc random alloys  \cite{Varvenne2016, Varvenne2016b}.

Within the VC model, the temperature-dependent yield 
stress is characterized by two thermal activation models specific
to certain temperature ranges.
In the low-temperature range, $\Delta \tau(T)$ is given by,
\begin{align}
\Delta \tau(T) = \tau_{y0} 
    \left( 1 
        - \left(
        \frac{k T}{E_{b}}
        \ln \frac{\dot{\varepsilon}_{0}}{\dot{\varepsilon}}
        \right)^{\frac{2}{3}}
    \right),
    \label{eq:tau_lowT}
\end{align}
whereas for higher temperatures,
\begin{align}
\Delta \tau(T) = \tau_{y0} \exp
    \left(
        -\frac{1}{0.51} \frac{k T}{E_{b}}
        \ln \frac{\dot{\varepsilon}_{0}}{\dot{\varepsilon}}
    \right).
    \label{eq:tau_T}
\end{align}

Both models give essentially equal results for moderate temperature range (100-400 K).
We employ, therefore, the high temperature model only for calculations
above 400 K, where some notable difference can be seen. 
In the above equations, $\tau_{y0}$ and $E_{b}$ are, respectively, the zero-temperature yield stress
and the activation barrier, given by the following expressions:

\begin{align}
    \tau_{y0} = & A_\tau \Big( \dfrac{\Gamma}{b^2} \Big)^{-\frac{1}{3}}
        \Big( \mu^V  \dfrac{1 + \nu^V}{1 - \nu^V}   \Big)^\frac{4}{3} \delta^\frac{4}{3},
    \label{eqn:vc_yield} \\
     \Delta E_{b} = & A_E \Big( \dfrac{\Gamma}{b^2} \Big)^{\frac{1}{3}}
        b^3 \Big( \mu^V  \dfrac{1 + \nu^V}{1 - \nu^V}   \Big)^\frac{2}{3} \delta^\frac{2}{3},
    \label{eqn:vc_energy}
\end{align}

with $b$, $\Gamma = \alpha \mu_{\langle 110 \rangle/\{111\}} b^2$ being
the Burger's vector and the dislocation line tension
($\alpha = 0.125$ is the edge dislocation line tension parameter
for fcc metals); $\mu_{\langle 110 \rangle/\{111\}}$ is the shear modulus for fcc slip on the
$\{111\}$ plane in the $\langle 110 \rangle$ direction; 
$\mu^V$ and $\nu^V$ are the Voigt average of shear modulus and Poisson’s ratio, respectively; $\delta$ is the misfit parameter
describing the collective effect of the individual solute misfit volumes.

The reference strain rate, $\dot{\varepsilon}_{0}$, is set to $10^{4}$ s$^{-1}$
as in previous works \cite{Yin2020a}. The actual strain rate, $\dot{\varepsilon}$, is 
always set to be the same as in the respective experiment. Prefactors $A_\tau = 0.04865 \, [ 1 - ( A - 1 ) / 40 ]$ 
and $A_E = 2.5785 \, [
1 - ( A - 1 ) / 80]$ are associated with a typical fcc dislocation core
structure consisting of two well-separated partial dislocations plus a small
correction for elastic anisotropy related to the Zener anisotropy $A = 2
C_{44}/(C_{11} - C_{12})$.

Note that apart from the explicit temperature dependence, 
all material parameters entering the expressions can be temperature dependent by themselves.
For instance, elastic moduli generally decrease with temperature, while the 
lattice constant (hence the volume and Burger's vector $b$) increase. 
This has often been neglected in earlier works but, as we will show in the next 
section, these additional temperature dependencies lead to an effective thermal behavior 
deviating from the naive Arrhenius law.

The central quantity of the VC model that introduces the chemical dependence
of the dislocation-solute interaction is the misfit parameter,
$\delta = \sqrt{\sum_i c_i (\Delta V_i)^2} / (3 V_{\textrm{alloy}} )$, where
$\Delta V_{i}$ are element specific misfit volumes,
\begin{equation}
    \Delta V_i = V_i - V_{\text{alloy}},
\end{equation}
with $V_{\text{alloy}}$ being the specific atomic volume of the alloy
and $V_i$ the \textit{apparent} (effective) volume of alloy component $i$.
The misfit volumes themselves are expressed in terms of concentration
derivatives of the alloy volume,
\begin{equation}
    \Delta V_{i} =  \frac{\partial V_{\text{alloy}}}{\partial c_i} - \sum_{j}
c_j  \dfrac{\partial V_{\text{alloy}}}{\partial c_j},
    \label{eqn:misfit_volumes}
\end{equation}
and the derivatives can be calculated by numerical differentiation. To get these derivatives in practice, 
we perform a series of CPA calculations for systems with
small deviations from the composition of the original alloy. The obtained volume points 
are then fitted to a linear function of concentration. Based on convergence tests (see Appendix A), 
we found that 4 volume points per alloy component,
with the concentration of the respective element 
varied by $-\delta$, $-\delta/2$, $\delta/2$, $\delta$ ($\delta=0.012$) with respect to 
the original alloy, is sufficient to get reliable derivatives $\frac{\partial V_{\text{alloy}}}{\partial c}$.
For an $N$-element alloy $4 \times N + 1$ calculations are performed 
including the original alloy composition. The linear elastic constants 
are obtained from volume-conserving monoclinic and orthorhombic distortions 
following the computational 
details described in Ref.~\onlinecite{Razumovskiy2011a,Razumovskiy2011b}.

Local lattice relaxations induced by the atomic size mismatch of the alloy components have found to be
essential for the stability and properties of HEA. As a single-site theory, CPA does not 
take into account local lattice distortions. However, previous 
supercell calculations of special quasi-random structures (SQS) \cite{Song2017,Tian2015} showed that the 
equilibrium parameters involved in the VC model are not significantly affected by local lattice distortions.
Upon including relaxation explicitly, the change in the lattice parameter for iron-group fcc HEA was 
found to be around 0.1-0.2 \%. Also the effect on the elastic constants was reported 
as insignificant in the case of similar sized constituents.

%
%

%
%

\section{Results}

We applied the methodology described in the previous sections
to calculate and analyze solid solution strengthening
of three alloys: NiCoCr, FeNiCoCr and, FeMnNiCoCr. All of them are well studied
both experimentally and theoretically, which enables us to carry out an extensive benchmarking.

\subsection{NiCoCr}

The solid solution alloy NiCoCr is a rare example
where a direct experimental measurement of misfit volumes was undertaken \cite{Yin2020b}. 
At the same time, DFT simulations accompanying experiment
in Ref.~\onlinecite{Yin2020b} showed significant deviations both in misfit
and apparent volumes. This system, thus, represents an ideal case for
testing our proposed methodology. 
First, we perform calculations at room temperature where
most of the SSS measurements are done. Since the equimolar NiCoCr 
alloy is already paramagnetic above 4 K \cite{Jin2016}, 
with its components exhibiting itinerant magnetism, 
this alloy is especially hard to model at this temperature. 
To treat the magnetic effects, we apply the LSF model described in Section~\ref{sec:finite-T}.
Following the recipes from \cite{Ruban2016}, the constant $a$ for the magnetic entropic
contributions is chosen for each component based on fixed spin-moment
calculations at zero temperature:

\begin{align}
\mathrm{Ni}: \quad & S_{\textrm{mag}} = 3 \log M, \\
\mathrm{Co}: \quad & S_{\textrm{mag}} = 2 \log M, \\
\mathrm{Cr}: \quad & S_{\textrm{mag}} = 3 \log M.
\end{align}

\begin{figure}
    \centering
    \includegraphics{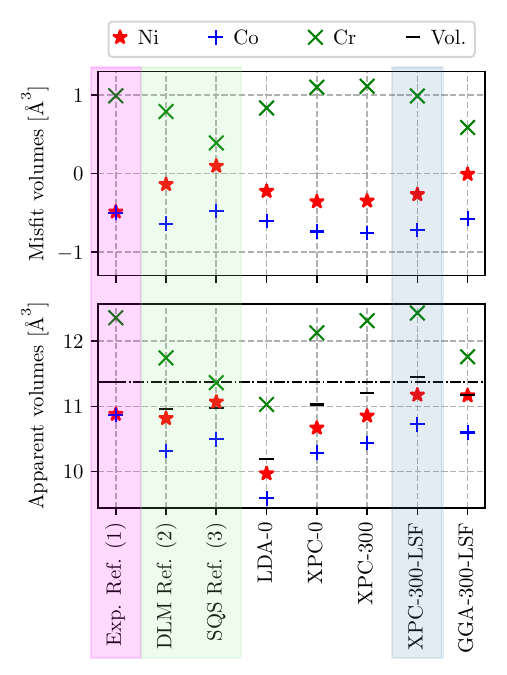}
    \caption{Comparison of experimentally measured and calculated apparent, 
    misfit ($\times$,$+$,$\star$) and equilibrium volumes ($-$). 
    Dash-dotted line indicates the experimental equilibrium volume.
    Ref (1): Volumes extracted from experimental measurements \cite{Yin2020b}.
    DLM Ref (2): CPA-DLM calculations \cite{Yin2020b}.
    SQS Ref (3): SQS calculations \cite{Yin2020b}.
    For our calculations, results of applying successively finite-temperature effects and exchange-correlation corrections are displayed.
    [LDA-0] and [XPC-0]: CPA-DLM with LDA calculation at 0 K without and with applying the exchange-correlation correction.
    [XPC-300]: Applying finite-temperature phonon contributions within
    the Debye-Gr\"uneisen model at 300 K.
    [XPC-300-LSF]: Additionally adding LSF.
    [GGA-300-LSF]: Conventional PBE calculation combined with DLM-LSF and
    phonon contributions at 300 K for comparison.
    }
    \label{fig:comparison_nicocr_results}
\end{figure}

We start by analyzing apparent and misfit volumes of individual
components (Ni, Co, and Cr), as they are the key quantities in the VC model.
The results are shown in \Cref{fig:comparison_nicocr_results},
where we compare them to the experimental measurements and earlier DFT calculations
from Ref.~\onlinecite{Yin2020b}.
Furthermore, we show how our methodology is getting more and more accurate by
gradually including finite-temperature effects and exchange-correlation corrections.

The biggest effect on the calculated equilibrium volume can be seen from using XC pressure correction, which is expected for LDA, since it is known to
underestimate the volume of 3$d$ metals considerably. It changes the equilibrium volume by
more than 8 \% (from $10.20$ $\textrm{\AA}^3$ [LDA-0] to $11.03$ $\textrm{\AA}^3$ [XPC-0]).
Including thermal electronic and phonon contributions leads to an additional
increase in volume to $11.20$ $\textrm{\AA}^3$.
Finally, taking also the LSF contributions into account, we obtain a volume of $11.44$ $\textrm{\AA}^3$ [XPC-300-LSF]. 

One can see that XC pressure correction and finite-temperature contributions (both with and
without LSF) lead to the equilibrium volume very similar to the experimental one.
At the same time, the agreement with experiment of the equilibrium volume calculated using the
common PBE functional (with all finite-temperature contributions included but without XPC)
seems to be as good as with our LDA-XPC approach
(PBE result: 11.18 \AA$^3$). However, previous PBE results without LSF (both DLM-CPA and SQS) from Ref.~\onlinecite{Yin2020b} 
underestimated the equilibrium volume of the NiCoCr alloy [by about 0.4 $\textrm{\AA}^3$, see Ref~(1), Ref~(2) in 
Fig.~\ref{fig:comparison_nicocr_results}]. 
The observed agreement can, thus, be attributed to the well-known
overestimation of magnetic moments by the PBE functional \cite{Park2015}, which, with the help of LSF, 
compensates for the error in the equilibrium volume. On the other hand, this
compensation effect is not consistent and fails to reproduce the misfit and apparent
volumes. In contrast to the equilibrium volume, calculated misfit and apparent volumes 
are more sensitive to
the accuracy of the underlying methodology because they depend on concentration derivatives
of the alloy volume. More precisely, any imbalance in the XC error produced
for individual alloy components can lead to enhanced errors in the misfit volumes.
Indeed, previous PBE calculations within DLM-CPA and SQS approaches from Ref.~\onlinecite{Yin2020b} showed considerable deviations of the misfit volumes from the ones obtained from experiment. In particular, the calculations underestimated the misfit volumes of Ni and Cr. 
Moreover, because of the their smaller equilibrium volume the apparent volumes of Co and Cr were lower than in experiment.
In contrast, the combination of XC pressure correction and finite-temperature contributions
results in a very good agreement of misfit and apparent volumes, as seen in Fig.~\ref{fig:comparison_nicocr_results} [XPC-300-LSF].

In particular, the misfit volume of Cr (0.99 \AA$^3$)
is practically the same as in experiment (0.99 \AA$^3$), while the results for Ni and Co (misfit volumes $-0.28$ and $-0.71$ \AA$^3$, respectively) are only slightly different from the experimental ones ($-0.49$ and $-0.50$ \AA$^3$, respectively).
Taking into account the good agreement of the equilibrium volume noted above,
the apparent volumes also turn out to be close to experiment. 
It is clear from Fig.~\ref{fig:comparison_nicocr_results} that the success in producing good misfit volumes in this particular alloy
can be largely attributed to XC pressure correction,
while the finite-temperature contributions are responsible for a more accurate equilibrium volume and hence apparent volumes of components. 

\begin{figure}
    \centering
    \includegraphics{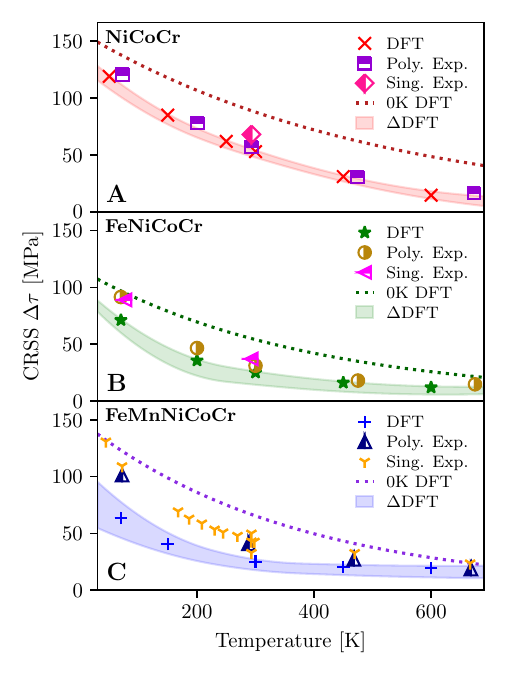}
    \caption{Comparison of CRSS $\Delta \tau$ from calculations and experiments versus temperature of NiCoCr (A), FeNiCoCr (B) and FeMnNiCoCr (C).
    [Poly. Exp.]: Experimental polycrystalline measurements with subtracted Hall-Petch contribution \cite{Wu2016, otto2013}.
    [Sing. Exp.]: Experimental single-crystalline measurements \cite{Uzer2018, Wu2015, okamoto2016, Abuzaid2017, Kawamura20211}.
    [DFT]: Theoretical CRSS with all material parameters calculated for the given temperature and uncertainties (filled curves) due to deviations from experiment. [DFT 0 K] (dashed lines): Theoretical CRSS with all material parameters calculated once for 0~K and then used for the whole temperature range.}
    \label{fig:temp_comp_3}
\end{figure}

\begin{figure}
    \centering
    \includegraphics{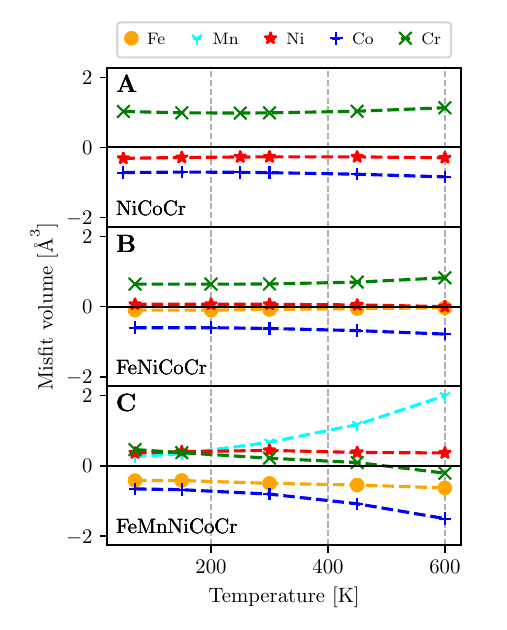}
    \caption{Comparison of misfit volumes of the individual components for the (A) NiCoCr, (B) FeNiCoCr and (C) FeMnNiCoCr alloy versus temperature.}
    \label{fig:temp_comp_misfit}
\end{figure}

\begin{figure}
    \centering
    \includegraphics{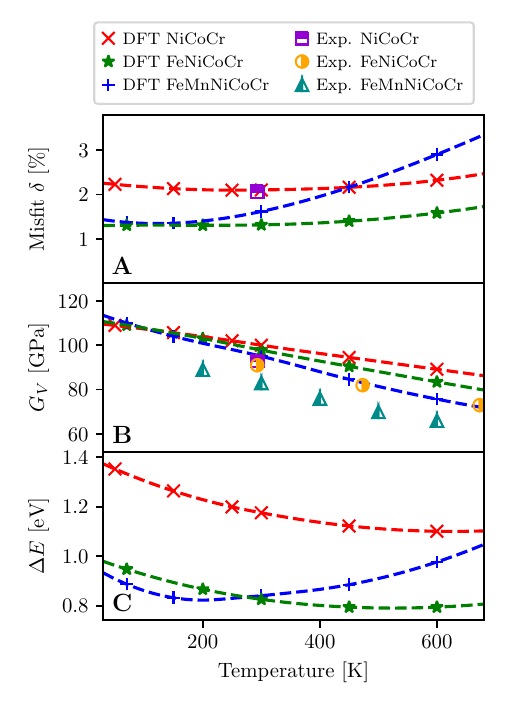}
    \caption{Comparison of the main parameters of the VC model from calculations and experiments versus temperature of (A) NiCoCr, (B) FeNiCoCr and (C) FeMnNiCoCr. %
    A: Calculated and experimental (\cite{Yin2020b}) misfit parameter $\delta$.
    B: Calculated Voigt-averaged shear modulus and experimental values of the shear modulus \cite{Jin2017,wu2014,Wagner2022,Laplanche2020,Haglund2015}.
    C: Calculated results for energy barrier $\Delta E_{b}$. }
    \label{fig:temp_comp_4}
\end{figure}

Next, we calculate how the SSS contribution to the yield stress evolves with temperature and compare it to experimental measurements. In Fig.~\ref{fig:temp_comp_3}, theoretical and experimental CRSS are compared for each of our alloys: [Poly. Exp.] and [Sing. Exp.] denote experimental values from
poly- and single-crystalline samples, respectively; 
[DFT] refers to the CRSS from our methodology with all material parameters being calculated for the respective temperature; 
{[0 K~DFT]} corresponds to our theoretical CRSS using the material parameters at 0~K, however, still considering the alloy to be paramagnetic and with the temperature dependence being determined solely by the Arrhenius-like expression from Eq.~\eqref{eq:tau_T} and Eq.~\eqref{eq:tau_lowT}. 
Furthermore, we also added an uncertainty estimate for the calculated CRSS ([$\Delta$DFT]),
which shows how differences between the predicted and experimental 
elastic constants, misfits and equilibrium volume influence the final result of the CRSS.
In order to get the experimental values of the CRSS from the polycrystalline measurements,
we are using the Taylor factor of $3.06$ to get the single-crystalline values and
then subtract the grain-size-dependent contribution obtained from the Hall-Petch fit. For the case of NiCoCr, polycrystalline tensile 
results from Ref.~\cite{wu2014} with a Hall-Petch contribution of 39 MPa from Ref.~\onlinecite{Yin2020b} were used.

For NiCoCr, Fig.~\ref{fig:temp_comp_3}A clearly shows 
that the full account of temperature dependence of the
material properties results not only in qualitative but also in a very good quantitative
agreement of the calculated CRSS with the experimental one, in a wide temperature range
from 77 K to 673 K. In contrast, using only 0~K parameters throughout the
whole temperature range leads to an obviously different
behavior at higher temperatures. According to Eq.~\ref{eq:tau_T} and \ref{eq:tau_lowT}, the temperature 
dependence of the CRSS is mostly determined by the energy barrier, 
$\Delta E_b$, which is itself a function of
the misfit volumes, elastic moduli,
and volume (Eq.~\ref{eqn:vc_energy}).
As can be seen in Fig.~\ref{fig:temp_comp_misfit}A, the misfit volumes of NiCoCr 
are only slightly affected by finite-temperature effects and stay almost constant up until 600 K. 
Together with the moderate increase of the equilibrium volume this also leads to an almost
constant misfit parameter $\delta$ (Fig.~\ref{fig:temp_comp_4}A). At the same time, 
the shear modulus $G_V$ decreases linearly with temperature, as can be inferred from Fig.~\ref{fig:temp_comp_4}B, where we also provide room-temperature experimental values
from Ref.~\onlinecite{Jin2017} for comparison. 
This softening, in turn, reduces the energy barrier, $\Delta E_{b}$, for dislocation glide, 
resulting in an additional reduction of the yield stress on top of the simple thermal activation process. Note also that the CRSS is reduced even further because of decreasing $\tau_{0y}$ whose $T$-dependence is similar to that of $\Delta E_{b}$. 
It is also worth noting that the uncertainties, which are mainly stemming from the overestimated shear modulus, are 
rather small, with values of around 10~MPa at low temperatures (below 100~K) and around 4~MPa above room temperature (RT).

Finally, we take a closer look at the CRSS at room temperature and compare our
calculated value,
$\Delta \tau = 63$ MPa to earlier works. Single-crystal tensile measurements in Ref.~\onlinecite{Uzer2018} yielded 69 MPa. 
A slightly different value of 63 MPa was found in Ref.~\onlinecite{Yin2020b} by extrapolating the Hall-Petch relation to infinite grain size and by dividing the results for polycrystalline samples by the Taylor factor. A comparable value of 59 MPa was calculated using the same VC model, with the parameters estimated from experimental data \cite{Varvenne2016}. At the same time, previous \textit{ab initio}
calculations seem to have failed to predict the yield stress of NiCoCr.
For instance, Liu~\textit{et. al.} \cite{Liu2019} used the Peierls model to estimate the zero-temperature CRSS to
be equal to 214 MPa. This significantly overestimates our 0 K prediction of 158 MPa. Finally, Yin~\textit{et. al.} \cite{Yin2020b} have obtained the CRSS at room temperature
of about 19 MPa, whereby they have applied the VC model, with the parameters calculated using the PBE functional
and a SQS setup for the alloy, finite-temperature contributions being neglected.

%
%

\subsection{FeNiCoCr}

{FeNiCoCr} is paramagnetic at room temperature \cite{Chaudhary2020} with a $T_c$ of around 85 K 
and can be produced as a single-phase fcc alloy \cite{Wu2016, Wu2015} without detectable long-range order \cite{Lucas2012}. 
However, its paramagnetic behavior differs from that of {NiCoCr}.
Specifically, unlike Ni, Co, Cr, the magnetic moment of Fe is strongly localized and
remains significant even at zero temperature \cite{Gambino2020,Huck1986}. We, therefore, use the LSF entropy term for localized moments: $S_{\textrm{mag}} = \log(1 + M)$ \cite{Ruban2016}.
To summarize, the following magnetic entropy contributions are chosen for each component:
\begin{align}
\mathrm{Fe}: \quad & S_{\textrm{mag}} = \log (1 + M), \notag \\
\mathrm{Ni}: \quad & S_{\textrm{mag}} = 3 \log M, \notag \\
\mathrm{Co}: \quad & S_{\textrm{mag}} = 2 \log M, \notag \\
\mathrm{Cr}: \quad & S_{\textrm{mag}} = 3 \log M.
\notag
\end{align}

As in the previous case, we have calculated all temperature points with
the DLM-LSF approach. This includes also the first experimental point at 77 K that lies below the
magnetic transition temperature. However, since this point is very close to $T_c = 85$ K, we expect a significant magnetic disorder, which is better described by DLM than the ferromagnetic state \cite{Ruban2008}.

The result for FeNiCoCr are presented in the middle panel of Fig.~\ref{fig:temp_comp_3}, where again, one can see a very good agreement for the calculated $\Delta \tau(T)$ [DFT] with experimental data from Ref.~\onlinecite{Wu2016,Wu2015}.
The difference between the full ([DFT]) and no ([0 K DFT]) temperature-dependence of material parameters is even more pronounced
than in the NiCoCr alloy, where the latter fails to reproduce the significant drop in strength below RT. It is also clear that the CRSS turns out to be considerably lower than in NiCoCr. To analyze
this difference, we examine individual contributions to the value of $\Delta \tau$.

The calculated Voigt-averaged shear modulus, $G_V$, is displayed in the middle panel of Fig.~\ref{fig:temp_comp_3} along
with experimental values from Ref.~\onlinecite{wu2014} for comparison.
The value of $G_V$ is similar to that of NiCoCr, with a slightly more enhanced softening with temperature. As already seen in the analysis of the uncertainties, small differences in the shear modulus between the two alloys cannot explain the much stronger difference in the CRSS. The equilibrium volume of {FeNiCoCr} at room temperature,
$11.32$ $\textrm{\AA}^3$ \cite{Lucas2012}, is practically 
the same as that of {NiCoCr}.
Our equilibrium volume of $11.44$ $\textrm{\AA}^3$ calculated at room temperature also agrees quite well with this experimental value.

Next, we consider the misfit volumes of alloy components (Fig.~\ref{fig:temp_comp_misfit}B) and the average misfit parameter $\delta$ (Fig.~\ref{fig:temp_comp_4}A).
Despite having almost the same equilibrium volume as NiCoCr, the misfit volumes of Ni, Co, Cr components in FeNiCoCr differ significantly from those in the ternary system. While the order has remained the same ($\Delta V_{Co} < \Delta V_{Ni} < \Delta V_{Cr}$),
the difference between the smallest (Co) and the largest (Cr) components in
FeNiCoCr at RT is 1.27 \AA$^3$, which is 25 \% smaller than in NiCoCr, where the difference between Co and Cr is 1.7 \AA$^3$ (see previous subsection). On the other hand, Ni and Fe have negligible contributions to the average misfit $\delta$ in FeNiCoCr. The net effect is the reduction of $\delta$ from 2.1 \% in NiCoCr to 1.6 \% in FeNiCoCr, which has the largest impact on the final values of $\Delta \tau(T)$.
The consequently lower energy barrier $\Delta E_b$ causes the CRSS 
to fall off significantly below RT (Fig.~\ref{fig:temp_comp_3}B). The $T$-dependence of the misfit volumes and hence of the average misfit $\delta$ in FeNiCoCr is practically absent below the RT and remains relatively weak at higher temperatures up until 673 K.

We can also compare our calculated CRSS to other experimental data measured at selected temperatures. Single-crystalline CRSS measurements from Wu \textit{et al.} \cite{Wu2015} yielded $\Delta \tau = 89$ MPa for 77 K and $39$ MPa for RT. For the same temperatures, the extrapolated polycrystalline CRSS from Ref.~\onlinecite{Wu2016} are $95$ and $30$ MPa, respectively.
This is to be compared to our calculated values for the same two temperatures: $77$ MPa and $25$ MPa, respectively.

%
%

\subsection{FeMnNiCoCr}

As a third example, we consider solid solutions of FeMnNiCoCr 
whose equimolar composition is known as
the Cantor alloy, for which we also study the effect of concentration variations. 
A wide range of compositions of FeMnNiCoCr systems were studied, with their microstructure
being claimed to be a single-phase fcc solid solution \cite{Bracq2019,Li2017,Laurent2016}.
This makes it especially interesting as a playground for property optimization. 
FeMnNiCoCr is paramagnetic at room temperature, with $T_c$ being around 38K \cite{Schneeweiss2017}.
The magnetic moments of Mn show localized behavior with a non-vanishing moment at zero temperature \cite{Gambino2020,Huck1986} similar to Fe.
The following magnetic entropy contributions are chosen for each component:
\begin{align}
\mathrm{Fe}: \quad & S_{\textrm{mag}} = \log (1 + M), \notag \\
\mathrm{Mn}: \quad & S_{\textrm{mag}} = \log (1 + M), \notag \\
\mathrm{Ni}: \quad & S_{\textrm{mag}} = 3 \log M, \notag \\
\mathrm{Co}: \quad & S_{\textrm{mag}} = 3 \log M, \notag \\
\mathrm{Cr}: \quad & S_{\textrm{mag}} = 3 \log M.
\notag
\end{align}

From the results presented in Fig.~\ref{fig:temp_comp_3}, one can immediately
see that the behavior of the Cantor alloy is generally more involved than in
the two previous cases. First of all, we see that the calculated CRSS has an appreciably
weaker temperature dependence than in NiCoCr and FeNiCoCr. The CRSS at RT is practically the same (26 MPa) 
as in FeNiCoCr despite being significantly smaller at lower temperatures.
The more so, our calculations seem to systematically
underestimate the CRSS compared to experiment \cite{otto2013, wu2014}.
Especially at lower temperatures, we see differences of more than 30~MPa.
Before discussing possible reasons for such a discrepancy,
let us analyze the behavior of the alloy in more details.

First, the addition of Mn results in the increase of the equilibrium volume
compared to NiCoCr and FeNiCoCr,
with the calculated volume being $11.62$ $\textrm{\AA}^3$ at RT, which nicely compares to the experimental value $11.56$ $\textrm{\AA}^3$ \cite{Huang2019}. 
On the other hand, the shear modulus is only marginally smaller than
in FeNiCoCr with the RT value for $G_V$ being 95 GPa.
Overall, the calculated shear modulus as function of temperature is somewhat
larger compared to experimental values from Ref.~\onlinecite{Wagner2022},
as can be seen in Fig.~\ref{fig:temp_comp_4}B. Temperature-induced softening 
is slightly more pronounced as compared to the previous two alloys.
The anti-invar behaviour poses a problem for the precise and unambiguous determination of the bulk modulus. 
As a result, a comparably large uncertainty in CRSS $\Delta \tau$ arises 
([$\Delta$DFT] in Fig.~\ref{fig:temp_comp_3}C).

The apparent volumes of Co, Cr and Fe at RT are similar to those of FeNiCoCr causing
the misfit volumes to be just shifted because of the difference in 
equilibrium volumes (Fig.~\ref{fig:temp_comp_misfit}C). 
Ni appears to be even larger than Cr in the Cantor alloy,
making it a significant strengthener in contrast to the other two alloys.
Compared to the previous two cases, the misfit volumes of the Cantor alloy exhibit
a significant temperature dependence above RT with Mn being most strongly affected.
At low temperatures, the previously found order of misfit volumes is retained, with Co being the smallest, followed by Fe, Ni and Cr. Mn lies in between Ni and Cr. 
At 600 K, however, Mn has by far the largest apparent volume and Ni and Cr are switching places,
with Ni becoming larger than Cr.
Starting from a value of 0.67 at RT, the misfit volume of Mn reaches a value of 2.00 at 600 K.
Combined with the decreasing misfit volumes of Cr and Co,
this leads to an effective doubling of misfit parameter $\delta$ from RT to 600~K.
This compensates for the effect of elastic softening, giving rise to 
both the energy barrier, $\Delta E_b$, and the zero-temperature yield strength, $\tau_{y0}$,
to gradually increase again starting from RT. This, finally, results in a plateau in the $T$-dependence of $\Delta \tau$ above RT. 

Next, we take a closer look at how our theoretical findings of CRSS compare 
with the available experimental data.
At RT, the yield stress ranges from 33-50 MPa \cite{okamoto2016,Abuzaid2017,Kawamura20211} (single-crystal results) to 
49-55 MPa \cite{George2019, otto2013} (extrapolated polycrystalline data), 
which is above our calculated value of 26~MPa. 
For 77~K we get 67~MPa compared to 105~MPa \cite{Kawamura20211} and 100~MPa \cite{otto2013},
which indicates a considerable underestimation,
even when the calculation uncertainty of $15$ MPa is taken into account.
The discrepancy can be attributed to a likely deviation of the structure
of the Cantor alloy from the idealized solid solution assumed in the calculations.
Specifically, the five-component system has a strong tendency to phase separation
observed in multiple experiments \cite{Laplanche2018c,Pickering2016,Otto2016,George2019}.
On top of that, even a single phase of a HEA can experience
partial ordering/clustering on individual sublattices, which would appear as
a homogeneous phase and could be detected only by experimental techniques
capable of resolving a homogeneous short-range order \cite{Schoenfeld2019}.

\begin{figure}
    \centering
    \includegraphics{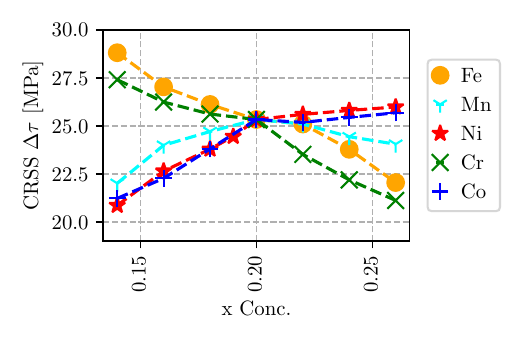}
    \caption{CRSS $\Delta \tau$ of nonequimolar FeMnNiCoCr Cantor alloys with varying concentrations. $x$ denotes the concentration of the component that is varied, while the remaining component concentrations are kept equimolar. }
    \label{fig:cantor_overview}
\end{figure}

To further investigate the complex interplay of the Cantor alloy components
and to understand its behavior better,
we also study the CRSS as a function of concentrations of components in the
vicinity of the equimolar composition.
In particular, we always vary one component concentration, while keeping the 
composition of the remaining system equimolar. 
Fig.~\ref{fig:cantor_overview} summarizes the CRSS for all component variations. 
Judging by the misfit volumes of the equimolar composition, one would 
expect that increasing the concentration of elements with large absolute misfit volumes, 
such as Mn and Cr, would also positively affect the CRSS. 
Surprisingly, the result for Mn and Cr is the opposite, while
increasing the concentrations of Co turns out to be marginally beneficial for the strength. 
By examining the behavior of individual contributions (see Appendix B),
we conclude that for Mn the outcome is related to considerable softening of
elastic constants, compensating the increasing misfit parameter, $\delta$.
At the same time, both the shear modulus and the misfit parameter are decreasing
with concentration of Cr, resulting in the reduced CRSS. 
On the other hand, lowering the Fe content increases $\delta$, while
reducing the elastic constants only moderately,
which gives rise to a positive impact on the CRSS.
This counter-intuitive behavior of the CRSS is the 
result of complex interactions between the components.
We will discuss these interactions in more details in the next section.

%
%

\section{Discussion}

Naturally, one would expect that a higher number of elements would enhance 
solid solution strengthening and ultimately lead to a higher yield stress. 
However, previous experimental work of tensile tests of polycrystalline samples,
yield strength measurements on single-crystal and hardness measurements show 
a different trend for the NiCoCr, FeNiCoCr and FeMnNiCoCr alloy \cite{GarciaFihlo022, Wu2016}.
Specifically, it is the ternary NiCoCr that has the highest SSS contribution,
while the five-component Cantor and the quaternary 
FeNiCoCr alloy have almost the same SSS, with the latter having the lowest value.
At the same time, specific volume and linear elastic constants are very
similar for these alloys,
which leaves the average misfit volume as a parameter of the VC model
mainly responsible for the differences in the SSS. 
Results presented in the previous section show that that our theoretical
findings confirm the trends observed in experiments.
In particular, NiCoCr has the largest SSS, with the four- and five-component
alloys having considerably lower effect of solutes on strengthening.

Modeling also confirms that the differences between the alloys
clearly correlate with the magnitude of the average misfit volumes.
We can therefore understand the mechanism of strengthening better
by analyzing the behavior of misfit volumes. Since direct measurements of misfit/apparent volumes are relatively difficult, 
they are often estimated from measured
volumes of a set of simpler alloys containing corresponding elements,
whereby a linear dependence of the alloy volume on concentrations
(Zen-Vegard's law \cite{Zen1956}) is assumed. 
Apparent volumes obtained in this way for iron-group 3d metals
usually follow the order similar to the well-known trend followed by the
corresponding elemental compounds \cite{Varvenne2016, Varvenne2017a, Bracq2019}.
Specifically, starting from Ni and going down the row, the volume steadily
increases, reaching its maximum at Mn and then slightly decreasing for Cr. 
However, our calculations suggest that the apparent volumes are strongly
system dependent and generally do no follow the naive elemental trend.
For instance, for all three alloys considered in this work, Ni is never the
smallest element and exhibits even a larger volume than that of Cr in
the five-component Cantor alloy (see Appendix B). 
Furthermore, previous works suggested that the average misfit, $\delta$, is
increasing for FeNiCoCr, NiCoCr, FeMnNiCoCr
(respectively, values of 1.72, 1.67, 1.85 \% were obtained in Ref.~\onlinecite{Varvenne2017a}
and 1.24, 1.07, 1.89 \% -- in Ref.~\onlinecite{Bracq2019}), while
our results clearly show that it is NiCoCr that has the largest average misfit volume among these three alloys (with the RT value of 2.1 \% compared with 1.31 and 1.61 \% for
FeNiCoCr and FeMnNiCoCr, respectively).

The failure of the simple Vegard's law in predicting the
misfit volumes can be attributed to two main phenomena.
First, the magnetic behavior of the 
iron-group elements is pretty complex and has a considerable influence
on the misfit volumes. In particular, Fe and especially Mn exhibit
strong magneto-volume coupling, which effectively makes their local
magnetic moments dependent on their respective apparent volumes.
Second, chemical and magnetic interactions between components in multi-component HEA
lead to appreciable deviations of the concentration
dependence of the volume from Vegard's law.

Because of magnetism, formulating a tractable general model for predicting
misfit volumes is a nontrivial task. Nevertheless, basic aspects of
interatomic interactions beyond Vegard's law can be rationalized
within a relatively simple framework that we will present below. 
The behavior of volumes in a multi-component alloy is best
inferred from the Gibbs free energy of a system, $G(T, P, \{c_i\})$,
which is a function of temperature, $T$, pressure, $P$, and concentrations of components, $c_i$.
Once the Gibbs free energy is known, the concentration-dependent volume of the alloy
for a given temperature (which we will omit for clarity) is readily obtained as
\begin{align}
    V(\{c_i \}) = \frac{\pd G(P, \{c_i\})}{\pd P},
\end{align}
which is then used to evaluate misfit volumes according to Eq.~\eqref{eqn:misfit_volumes}.

A standard way to analyze the free energy would be to use the cluster expansion, which
for a completely random alloy of $n$ components can be written as a series over
powers of concentrations,
\begin{align}
    G(P, \{c_i\}) = \sum_{k = 1}^{\infty} \sum_{p_1 p_2 \dots p_n}
        g^{(k)}_{p_1 p_2 \dots p_n}(P) c_1^{p_1} c_2^{p_2} \dots c_n^{p_n},
\end{align}
where $g^{(k)}_{p_1 p_2 \dots p_n}(P)$ are effective intercomponent interactions of
$k$-th order, with $p_1 + p_2 + \dots + p_n = k$.
The interactions depend only on the pressure and component types and can formally
be obtained as a sum over corresponding cluster interactions. 
However, this series is generally slowly convergent in
the order of interactions, making it impractical for a model description.
The situation gets worse in magnetic alloys, where magnetic interactions
and the concentration dependence of local magnetic moments of components render
the above series even slower convergent (or not convergent at all).

A much more compact description can be obtained if we introduce two major assumptions:
(i) effective interactions $g^{(k)}$ can be made explicitly dependent on the local magnetic
moments of the components, $m_i$; (ii) given the local magnetic moments, effective
interactions depend on the component types only through the average number of
\textit{d} electrons per $k$-site cluster, $N^{(k)} = (p_1 N_1 + \dots + p_k N_k) / k$,
with $N_i$ being the number of \textit{d} electrons for component $i$. This approximation
is known as a virtual bond approximation (VBA) \cite{Ruban1998} and its idea is to
reduce complex intercomponent interactions to simpler universal (virtual bond) functions
of the average valence in the spirit of the Pettifor theory of bonding in transition metals \cite{Pettifor1995}. Virtual bond parameters can also be loosely connected to 
tight-binding bond order parameters, which reflect the dependence of the bond
energy on electron filling \cite{Pettifor1987}.
Another similar approach is a model proposed in Ref.~\onlinecite{Moreen1971}, which
expresses the alloy lattice parameter in terms of a second-order polynomial
in concentrations. Within the VBA model, Gibbs free energy and the equilibrium volume of an alloy
can be written as
\begin{align}
    G(P, \{c_i\}) = & \sum_{i} c_i g^{(1)}(P, N_i, m_i) + \notag \\
       & + \sum_{i j} c_i c_j g^{(2)}(P, N_{ij}, m_i, m_j) + \dots \\
    V(\{c_i \}) = & \sum_i c_i \, v^{(1)} \big( N_i, m_i \big) + \notag \\
       & + \sum_{i j} c_i \, c_j \, v^{(2)} \big( N_{ij}, m_i, m_j \big) + \dots,
    \label{eqn:vba}
\end{align}
where $v^{(k)} = \pd g^{(k)}(P = 0) / \pd P$, $N_{ij} = (N_i + N_j) / 2$.
Note that $i$, $j$ can stand for the same element.
In particular, the equilibrium volume of an element $i$ will be given
by $V_i = v^{(1)}[N_i] + v^{(2)}[N_{ii}]$ with $N_{ii} \equiv N_i$.
The expansion can be written up to an arbitrary order but we will
limit ourselves to the second order.

\begin{figure}
    \centering
    \includegraphics{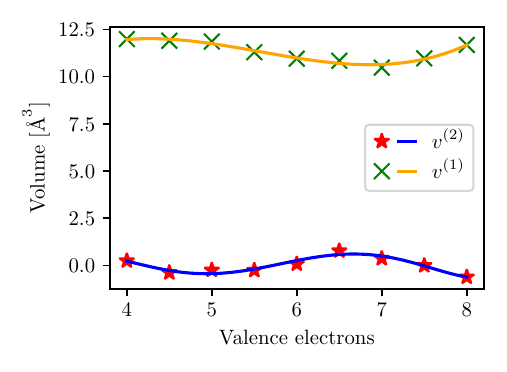}
    \caption{VBA parameters $v^{(1)}$ and $v^{(2)}$ as functions of the average
    number of $d$-valence electrons. Discrete values ($\times$,$*$) were obtained from equilibrium volumes of elemental
    compounds and binary alloys as described in the main text.
    Solid lines are polynomial fits (third order for $v^{(1)}$ and fifth order for $v^{(2)}$).}
    \label{fig:model_params}
\end{figure}

To determine the unknown interaction parameters, $v^{(k)}$, we perform a series
of calculations of equilibrium volume for equimolar binary alloys containing all
possible combinations of 5 elements, Cr, Mn, Fe, Ni, Co (10 binary alloys in total). 
Combined with 5 elemental compounds, this gives 15 reference systems in total.
Importantly, the calculations are performed with the local moments, $m_i$,
fixed to that of the target alloy. Also their structure is considered to be the same as the target. 
In practice, we fix the local moments to those of the Cantor alloy, since
Since the model is linear in interaction parameters, $v^{(k)}$, one way to find them
would be to perform a linear regression on the entire set of reference system.
However, one can simplify the fitting procedure by noticing that the model with
second-order terms ($v^{(2)}$) omitted is equivalent to
Vegard's law, with $v^{(1)}$ just being the simple elemental equilibrium volume.
Any deviation from Vegard's law, $\Delta V = V_{\textrm{alloy}} - \Av{V}$,
will thus, be determined solely by interaction terms, $v^{(2)}$.
In more details, for an equimolar binary AB the deviation,
$\Delta V_{AB} = V_{AB} - \Av{V}_{AB}$, is given by 
\begin{align}
    \Delta V_{AB}  =& \dfrac{1}{2} v^{(2)} \left[ N_{AB} \right] 
                   - \dfrac{1}{4} v^{(2)} \left[ N_{A} \right] 
                   - \dfrac{1}{4} v^{(2)} \left[ N_{B} \right],
   \label{eqn:vegards}
\end{align}
which results in a (generally overdetermined) system of equations in terms of
parameters $v^{(2)}$, with $\Delta V_{AB}$ obtained from the calculated
equilibrium volumes of the binary alloys. Subsequently, $v^{(1)}$ is obtained trivially from the equilibrium volumes of reference
systems using Eq.~\eqref{eqn:vba}. If the model consistently describes the reference systems we expect parameters
$v^{(k)}$ to come out as smooth functions of the valence. This is indeed the case, as
one can see in Fig.~\ref{fig:model_params}, where we show that the obtained parameter
values can be fitted with low-order polynomials. It is also clear that although the
volume is dominated by the first-order parameter, $v^{(1)}$, the variation of $v^{(2)}$
is significant and it can result in nontrivial contributions to the equilibrium volume
and its concentration derivatives for multi-component alloys. 

\begin{figure}
    \centering
    \includegraphics{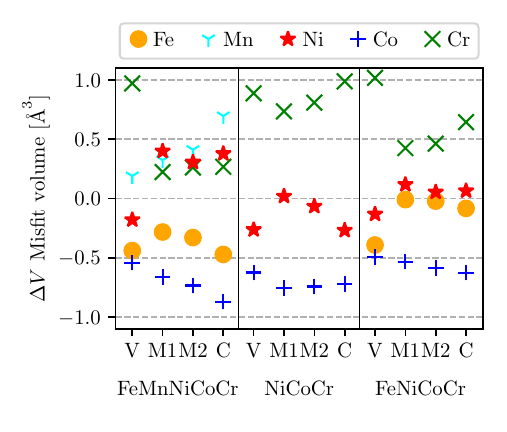}
    \caption{Comparison of the misfit volumes obtained from Vegard's law (V), from
    virtual bond approximation with nominal valences (M1), with charge transfers 
    taken into account (M2), and calculated directly (C).
    }
    \label{fig:model_hea}
\end{figure}

Misfit volumes derived from Vegard's law and the VBA model for the three alloys,
NiCoCr, FeNiCoCr, FeMnNiCoCr, are displayed in Fig.~\ref{fig:model_hea},
where they are compared to the calculated values (denoted by ``C'').
To get the model estimates we use the polynomial fits of
$v^{(1)}$ and $v^{(2)}$ as functions of the valence, which allows us to evaluate these
parameters at an arbitrary value of the \textit{d}-electron filling.
Furthermore, we use either nominal
valences of elements (``M1'' in the figure) or the actual fillings including charge
transfers (``M2'' in the figure) obtained from the calculations. 
From Fig.~\ref{fig:model_hea} one can see that the naive Vegard's law (``V'')
works fairly well only for NiCoCr, while failing completely to reproduce the signs and
relative sizes of calculated misfit volumes in FeNiCoCr and FeMnNiCoCr. In the latter
case, even the order of elements turns out to be wrong with a strongly overestimated
value for Cr and underestimated values for Ni and Mn. The reason for this failure is
clear: Since the volumes of the three alloys are very similar, the obtained
misfit volumes simply follow the trend for the volumes of 3$d$ elements.
However, we have seen already that the misfit volumes in the four- and five-component
alloys deviate considerably from this trend. 

In contrast, misfit volumes from the VBA (M1 and M2) exhibit much better agreement
with calculations. In particular, the signs are always correct and values are mostly
in the right ballpark. The order of the elements is also generally correct except for the
significant underestimation of Mn misfit volume in the Cantor alloy by the simpler
version (M1) of the model. However, this error is mitigated if one takes charge transfer
into account (M2), which gives an overall better agreement in all cases.
The considerable improvement of the VBA misfit volumes over those from Vegard's
law signify the importance of including intercomponent interactions into the model.

\begin{figure}
    \centering
    \includegraphics{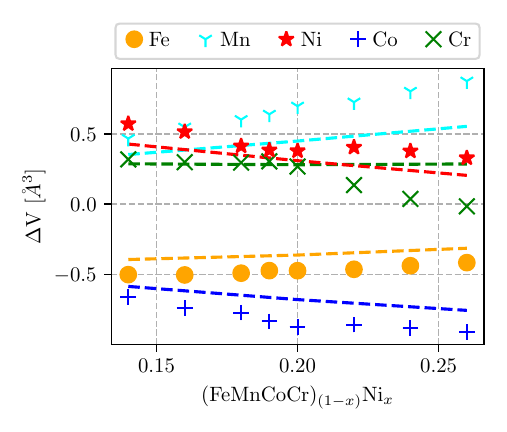}
    \caption{Comparison of the directly calculated and model predictions of misfit volume of the Cantor alloy for concentration variations of Ni.}
    \label{fig:model_concNi}
\end{figure}

Another important feature of the VBA model is that it can be used to predict the
evolution of the misfit volumes when the composition of a HEA is varied away from
the equimolar point. An example is shown in Fig.~\ref{fig:model_concNi}, where
the VBA misfit volumes are compared to the calculated ones as functions of the
concentration of Ni in the Cantor alloy while keeping the other components equimolar. As 
expected from the second-order model,
it can result only in a linear dependence on the concentration.
The concentration dependence of the local magnetic moments is ignored here.
Including changes in the \textit{d} fillings with concentration have a weak effect.
Nevertheless, most of the trends are correctly reproduced.
In contrast, Vegard's law would give values shifting uniformly in the same direction
with composition only because of the variation of the equilibrium volume itself.

The presented model could, in principle, be used for predicting misfit volumes
of HEA based on experimental data on binary and ternary subsystems.
However, the inference of parameters $v^{(k)}$ for iron-group HEA is complicated by
the complex magnetic behavior of elements, especially Mn and Fe. In particular,
one would need an additional model to predict the local moments of components and
to take into account their effect on $v^{(k)}$. These issues will be addressed in
future publications.

%
%

\section{Conclusions}

We have shown that a quite reliable description of such a complex alloy
property as the critical resolved shear stress is possible using DFT combined with CPA, provided that
a) the equilibrium volume is estimated correctly,
b) important finite-temperature effects, such as the thermal expansion
and spin fluctuations, are taken into account.
An accurate value of the equilibrium volume is obtained using an element-specific
XC pressure correction, which is a key feature of our approach.
The improved equilibrium volume also leads to significantly more
accurate results for other equilibrium properties, such as elastic constants, magnetic moments, etc.
Furthermore, temperature effects mediated by phonons and magnetic fluctuations
are taken into account, which allows us to predict the behavior of alloys at finite temperatures.

We validate the computational approach against available experimental data
for a series of alloys. In particular, we have obtained a very
good estimate of the misfit volumes in the case of NiCoCr, where we can juxtapose
them with the corresponding experimental data. 
The methodology is applied to modeling of solid solution strengthening.
Specifically, we have calculated alloy-specific parameters (lattice constant, misfit
parameter, elastic moduli) at respective temperatures and employed the Varvenne-Curtin model to evaluate
the temperature-dependent CRSS for three alloys (NiCoCr, FeNiCoCr, FeMnNiCoCr)
for which corresponding experimental data is available.
We get a good quantitative agreement for the temperature-dependent CRSS for NiCoCr and
FeNiCoCr. For the five-component system we get somewhat underestimated values, which we
attribute to the complexity of the alloy structure. 
We have shown that the trends in the strength of the three alloys in question are
mostly determined by the misfit parameter, $\delta$. By examining the contributions of
individual elements into the average misfit parameter, we conclude that misfit volumes
are subject to intricate intercomponent interactions and can behave in a non-intuitive
way as functions of concentrations.

Finally, to examine complex interaction effects we propose a simple model
describing local bond strength between individual elements.
The model yields qualitatively good estimates of misfit volumes of individual
components, describing the differences in the behavior of the three considered
alloys. Moreover, the model is capable of capturing trends in the evolution
of misfit volumes as functions of concentrations in the FeMnNiCoCr alloy.
We argue that the model could potentially be used in conjunction with available
experimental data on 3$d$-metal alloys, provided that the description of magnetism is improved.

\acknowledgments{
We are grateful to V. Razumovskiy for discussions.
This work was supported by the Forschungsf\"orderungsgesellschaft (FFG)
Project No. 878968 ``ADAMANT'', Austrian Science Fond (FWF) Project No. P33491-N ``ReCALL'',
and COMET program IC-MPPE (Project No 859480).
This program is supported by the Austrian Federal Ministries for Climate Action, Environment, Energy, Mobility, Innovation and Technology (BMK) and for Digital and Economic Affairs (BMDW), represented by the Austrian research funding association (FFG), and the federal states of Styria, Upper Austria and Tyrol. All calculations in this work have been done using Vienna Scientific Cluster (VSC-3).
}

%
%


%

%
%
\clearpage
\appendix

\section{Convergence of misfit volumes}

\FloatBarrier

To see how sensitive our approach is to the variation of the mesh resolution, we also perform convergence tests for the concentration steps used to obtain the misfit volumes (a figure is enclosed in the appendix). We found that convergence is achieved already with a concentration step as large as 5 at. \% validating the mesh used in experiment from Ref.~\onlinecite{Yin2020b}. Specifically, NiCoCr exhibits a large linear concentration dependence of the equilibrium volumes. Often, the apparent and misfit volumes are calculated under the assumption of linearity of the average properties with respect to composition within a range of 5-10 at. \% (local Vegard's law) \cite{Yin2020a,Razumovskiy2019,Varvenne2017a,Bracq2019}. Our findings generally confirm this assumption, but also demonstrate that these previous studies were choosing meshes that are already boarder cases were non-linearity sets in. It is worth noting, that misfit volumes not only affect the yield stress at 0 K but also its temperature dependence, which can explain, for instance, the underestimated SSS at elevated temperatures in Ref.~\onlinecite{Razumovskiy2019}, resulting from underestimated misfit volumes.

\begin{figure}
    \centering
    \includegraphics{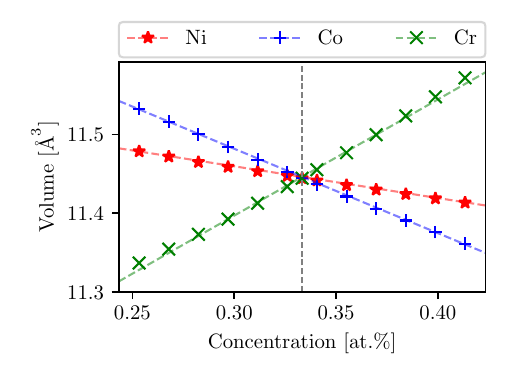}
    \caption{Atomic volumes calculated for different concentration variation of the NiCoCr alloy used for obtaining misfit volumes. The label, Ni (red), Co (blue) or Cr (green) refers to the elemental component which concentration is varied, while the other two component concentration ratios are kept constant. The markers refer to volumes obtained with our methodology. The lines correspond to a linear hyperplane fitted with the volumes. Ni and Co show a linear trend over a the large concentration range. Cr is starting to deviate for concentration changes larger than 3 at.\%.}
    \label{fig:convergence}
\end{figure}

As aforementioned, the misfit volumes require the determination of the concentration dependency of the equilibrium volumes of the respective alloy. For the NiCoCr linear-dependency of the concentration is present for a large concentration range. Fig.~\ref{fig:convergence} depicts the changes on the volume on varying always one component while ratio of the other two are kept constant. Changing the Ni and Co lead to a linear behavior over the whole range, as can be seen by comparing to lines representing a linear fit. In contrast, Cr shows deviations starting from concentration changes larger than 5 at.\%. Overall the deviation are small for this type of alloy.

\FloatBarrier

\section{Concentration variation around the equimolar composition of Cantor alloy}

\begin{figure}
    \centering
    \includegraphics{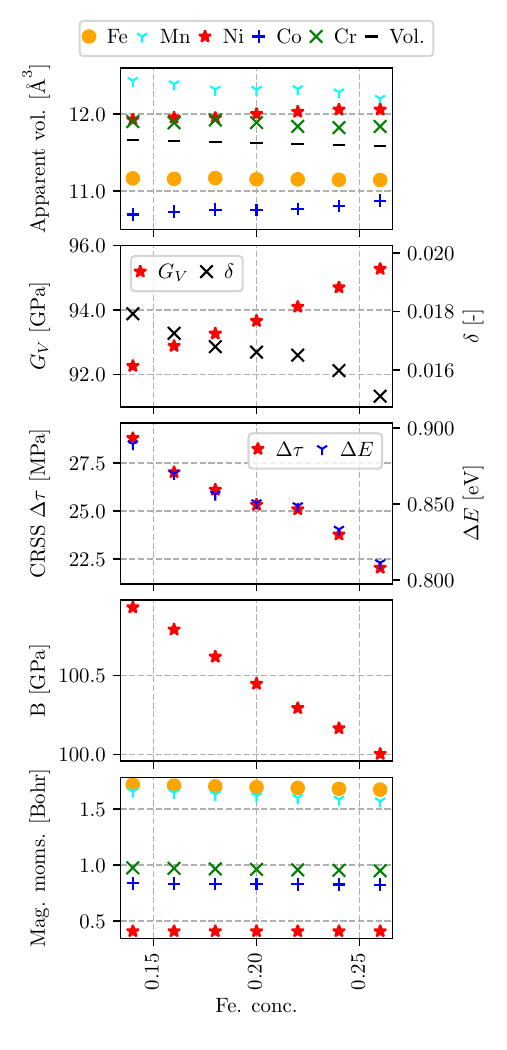}
    \caption{Effects on properties upon changing the concentration $x$ in $\textrm{Fe}_{x}\textrm{Mn}_{({1-x})/4}\textrm{Ni}_{({1-x})/4}\textrm{Cr}_{({1-x})/4}\textrm{Co}_{({1-x})/4}$.}
    \label{fig:cantor_1}
\end{figure}

\begin{figure}
    \centering
    \includegraphics{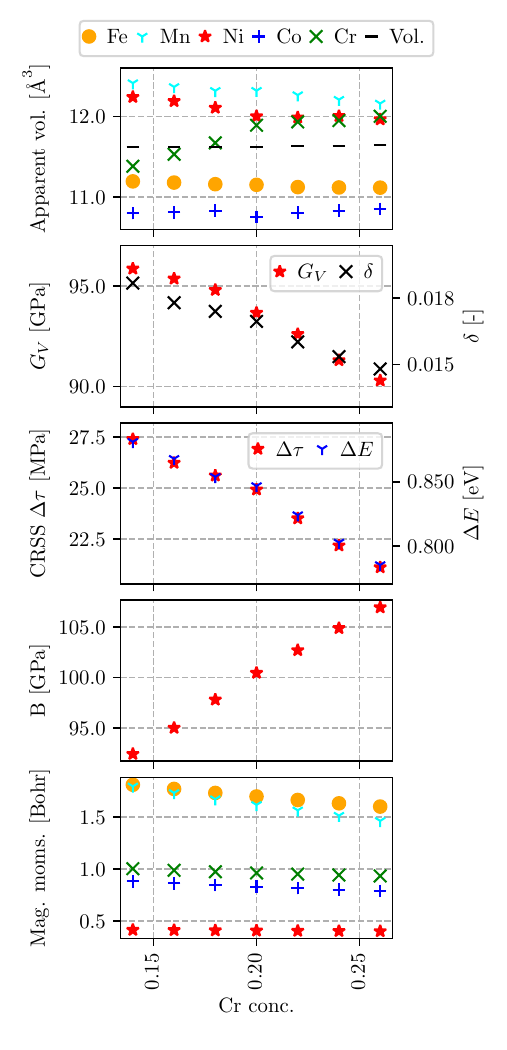}
    \caption{Effects on properties upon changing the concentration $x$ in $\textrm{Fe}_{(1-x)/4}\textrm{Mn}_{({1-x})/4}\textrm{Ni}_{({1-x})/4}\textrm{Cr}_{x}\textrm{Co}_{({1-x})/4}$.}
    \label{fig:cantor_2}
\end{figure}

\begin{figure}
    \centering
    \includegraphics{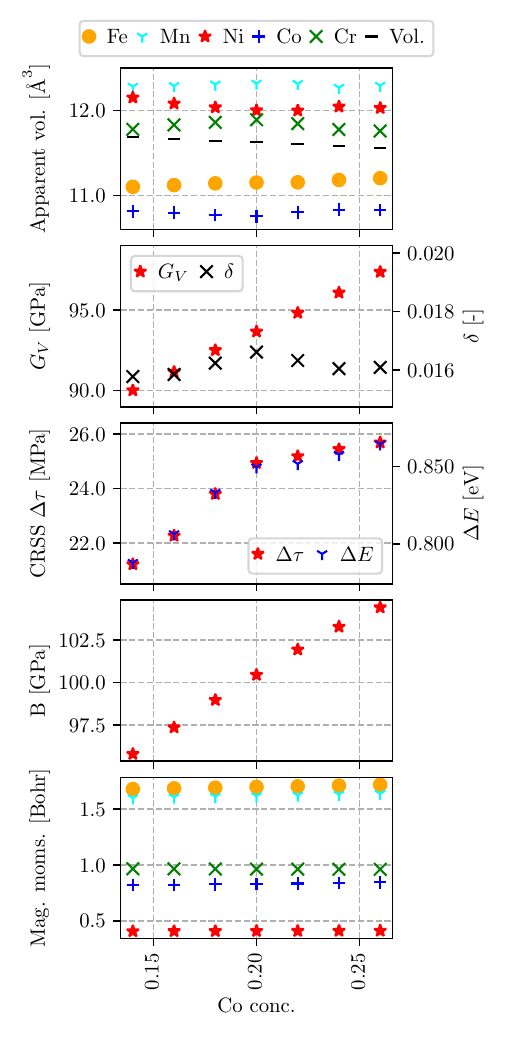}
    \caption{Effects on properties upon changing the concentration $x$ in $\textrm{Fe}_{(1-x)/4}\textrm{Mn}_{({1-x})/4}\textrm{Ni}_{({1-x})/4}\textrm{Cr}_{({1-x})/4}\textrm{Co}_{x}$.}
    \label{fig:cantor_3}    
\end{figure}

\begin{figure}
    \centering
    \includegraphics{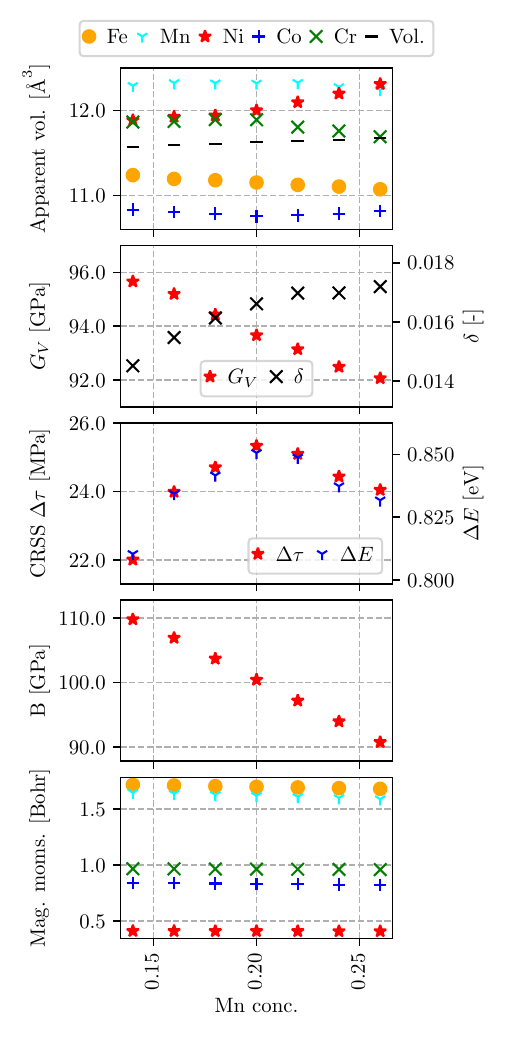}
    \caption{Effects on properties upon changing the concentration $x$ in $\textrm{Fe}_{(1-x)/4}\textrm{Mn}_{x}\textrm{Ni}_{({1-x})/4}\textrm{Cr}_{({1-x})/4}\textrm{Co}_{({1-x})/4}$.}
    \label{fig:cantor_4}
\end{figure}

\begin{figure}
    \centering
    \includegraphics{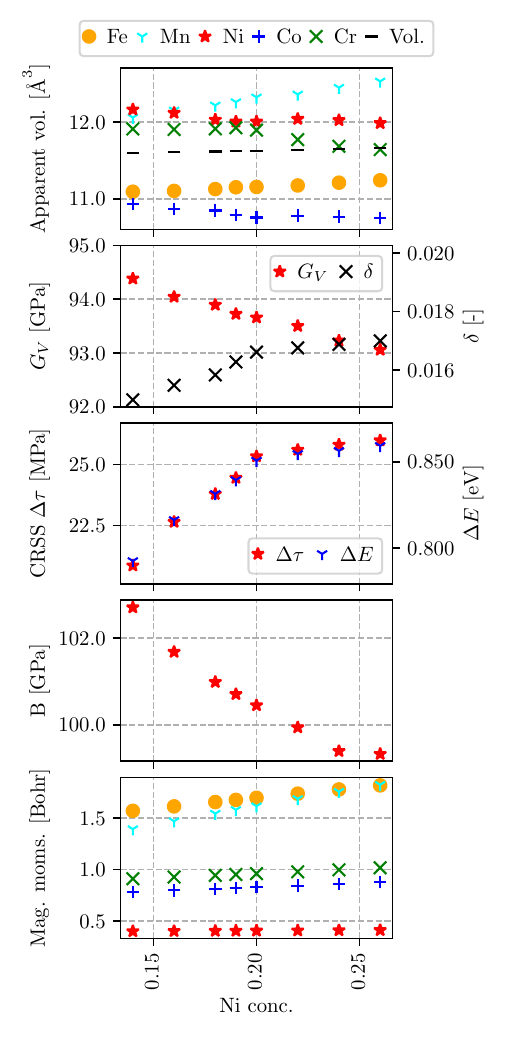}
    \caption{Effects on properties upon changing the concentration $x$ in $\textrm{Fe}_{(1-x)/4}\textrm{Mn}_{({1-x})/4}\textrm{Ni}_{x}\textrm{Cr}_{({1-x})/4}\textrm{Co}_{({1-x})/4}$.}
    \label{fig:cantor_5}
\end{figure}

\Cref{fig:cantor_1,fig:cantor_2,fig:cantor_3,fig:cantor_4,fig:cantor_5} give an overview of the influence of concentration changes on the elastic properties and apparent volumes.
In all cases, one alloy component is changed with respect to the equimolar Cantor while the ratios of the remaining components are kept constant. The following properties can be seen in the overview figures: (1) Apparent volume and the corresponding equilibrium volume of the alloy, (2) Voigt averages shear modulus $G_V$ and the average misfit delta $\delta$, (3) critical resolved shear stress at 300 K and the energy barrier, (4) bulk modulus, (4) magnetic moments of the components at the equilibrium volume.

Several interesting observations can be made in the figures. Changes in the
apparent volumes of Fe and Co are relatively weak compared to other elements,
except for the case of the varying Ni concentration, where all elements
are effected. Another counter-intuitive aspect is that in many cases
(except for Cr) the variation of an element concentration has less impact
on its own apparent volume than on the apparent volume of other elements.
For instance, varying Fe concentration changes the apparent volumes of
Mn, Ni, and Co, but its own apparent volume remains practically constant.

One could expect that most of these variations are related to the
changes in the local magnetic moment. However, the figures (bottom panels)
show that the magnetic moments evolve significantly only in the case of
varying Ni and Cr concentrations. And even in these cases, it is only
the variations of concentrations of Fe and Mn that can be attributed to
the well-known magneto-volume coupling inherent to these elements in the fcc lattice.
In all other cases, the results can be explained only taking into account
intercomponent (pair or higher order) interactions.

\FloatBarrier
\clearpage

\end{document}